\documentclass[12pt,preprint]{aastex}
\shortauthors{Li et al.} \shorttitle{UV-dissipating disk-jet
system in Rosette}

\received{2005 Nov 19}
\begin{document}

\title{Resolving the nature of the Rosette HH1 jet facing strong UV dissipation}
\author{Jin Zeng Li\altaffilmark{1}, You-Hua Chu\altaffilmark{2,3}, Robert A. Gruendl\altaffilmark{2}, John
Bally\altaffilmark{4,5} and Wei Su\altaffilmark{1}}
\altaffiltext{1}{National Astronomical Observatories, Chinese
Academy of Sciences, Beijing 100012, China; ljz@bao.ac.cn}
\altaffiltext{2}{Department of Astronomy, University of Illinois,
             Urbana-Champaign, IL, 61801}
\altaffiltext{3}{Visiting Astronomer, Cerro Tololo Inter-American Observatory}
\altaffiltext{4}{Center for Astrophysics and Space Astronomy, and Department
             of Astrophysical and Planetary Sciences, University of Colorado,
             Campus Box 389, Boulder, CO 80309-0389}
\altaffiltext{5}{Visiting Astronomer, Kitt Peak National Observatory}

\begin{abstract}
The Rosette HH1 jet is a collimated flow immersed in the strong
UV radiation field of the Rosette Nebula.
We investigate the physical properties of the Rosette HH1 jet using
high-quality narrow-band images and high-dispersion spectroscopy.
The new images show that the axis of the jet is not precisely
aligned with the star near the base of the jet.
The high resolution of the spectra allows us to accurately determine
the contributions from the \ion{H}{2} region, jet, and star.
The appoaching and receding sides of the expanding shell
of the Rosette Nebula are at heliocentric velocities of
13 and 40 km~s$^{-1}$, while the jet reaches a maximum
velocity offset at a heliocentric velocity of $-$30 km~s$^{-1}$.
The [\ion{S}{2}] doublet ratios indicate an electron density
of $\sim$1000 cm$^{-3}$ in the jet and $\le$100 cm$^{-3}$ in
the \ion{H}{2} region.
With a careful subtraction of the nebular
and jet components, we find the stellar H$\alpha$ line is
dominated by a broad absorption profile with little or no
emission component, indicating a lack of substantial
circumstellar material.
The circumstellar material has most likely been photo-evaporated
by the strong UV radiation field in the Rosette Nebula.
The evaporation time scale is 10$^3$ -- 10$^4$ yr.
The Rosette HH1 jet source provides evidence for an
accelerated evolution from a CTTS to a WTTS due to the
strong UV radiation field; therefore, both CTTSs and WTTSs
can be spatially mixed in regions with massive star formation.

\end{abstract}


\section{Introduction}

Herbig-Haro (HH) objects immersed in an ultraviolet (UV) radiation
field can be photoionized externally \citep{Retal98}.
The photoionized jets/outflows of HH objects become optically visible,
and thus their detailed physical properties can be studied.
Such photoionized HH jet systems have been identified in the Orion
Nebula and in the reflection nebula NGC 1333 \citep{Betal00,BR01}.
Recently, two such photoionized jet systems, the Rosette HH1 and HH2
jets, were discovered within the central cavity of the Rosette Nebula
\citep{Li03,LR04}.  The Rosette Nebula is a spectacular \ion{H}{2}
region excavated by strong stellar winds from dozens of OB stars at
the center of the young open cluster NGC 2244, the primary
component of a possible twin cluster recently identified using the
2MASS (Two Micron All Sky Survey) database \citep{Li05}.  At a
distance of $\sim$ 1.39 kpc \citep{Hetal00}, this emerging young open
cluster is found to have a main sequence turn-off age of about 1.9 Myr
\citep{PS02}.

The photoionized jets discovered in the Rosette Nebula
\citep{Li03,LR04} and their counterparts found in the vicinity of
$\sigma$ Orionis \citep{Retal98} are both bathed in harsh UV
radiation from massive OB stars within a few parsecs, and thus
share many similar properties consistent with an irradiated origin
of the jet systems: (1) Their jet-driving sources are visible and
show spectral characteristics of T Tauri stars. (2) These sources
were not detected by \emph{IRAS} (\emph{Infrared Astronomical
Satellite}), indicating a lack of circumstellar material such as
extended disks and/or envelopes. (3) The jets show
[\ion{S}{2}]/H$\alpha$ line ratio decreasing from the base
outward, indicating that the dominant excitation mechanism changes
from shocks at the base to photoionization at the end of the jet.
(4) The jet systems all have a highly asymmetric or even unipolar
morphology, indicating perhaps different jet forming conditions in
the launch and collimation regions.

The Rosette HH jets show subtle differences from other externally
photoionized HH jets because of different degrees of hardness in
the UV radiation field or strength of fast stellar winds.
Both the Rosette HH1 and HH2 jets show high excitation \citep{LR04},
as the Rosette Nebula contains an O4 star and an O5 star \citep{PTW87}.
In the Orion Nebula, HH jets with [\ion{O}{3}] emission are found only
within 30\arcsec, or $\sim$0.06 pc, from $\theta^1$\,Ori\,C, an O4-6
star, the earliest O star in the Orion Nebula \citep{Metal04}.
The high excitation of these HH jets results from both the harsh UV
radiation and strong fast stellar wind of $\theta^1$\,Ori\,C
\citep{Betal98}.

\citet{LR04} propose that the Rosette jets provide evidence for
efficient dissipation of circumstellar disks and envelopes in
the close vicinity of massive OB stars.  This UV dissipation of
pre-existing protostellar systems may lead to the formation of
isolated brown dwarfs (BDs) and free-floating giant planets.
Such a formation mechanism for single sub-stellar objects has
indeed been shown to be effective by theoretical studies \citep{WZ04}.

It is therefore important to explore the nature of jet formation and
disk dissipation of low-mass YSOs in close vicinity of massive
ionizing OB stars, as the occurrence of such OB clusters and
associations is common in the Galaxy, and the solar system may
have been formed in such environments \citep{LTF06}.
Furthermore, there has been
an on-going debate whether weak-lined T\,Tauri stars (WTTSs) evolve
from classical T\,Tauri stars (CTTSs) through gradual dissipation
of circumstellar material, or WTTSs are formed through rapid disk
dissipation due to external forces after the formation of the protostar.
A detailed study of the Rosette jet systems may provide insight on the
rapid evolution of CTTSs to WTTSs due to external photoionization of
their protostellar disks in massive star forming regions. WTTSs formed
in this way have indistinguishable evolutionary ages from those of CTTS
that originated from the same episode of star formation.

\citet{Metal05} presented a kinematical study of the Rosette HH1
jet and confirmed the jet nature of the system.  Here we
investigate in detail the physical nature of the jet system using
high-resolution imaging and echelle spectroscopy, as well as data
from a simultaneous photometric and spectroscopic monitoring of
the jet-driving source.

\section{Observations and Data Reduction}

\subsection{Narrow-band Imaging}

Narrow-band H$\alpha$ images of the Rosette Nebula were obtained
with the 8k$\times$8k MOSAIC CCD camera on the Mayall 4~m
telescope at the Kitt Peak National Observatory on 2001 October
13.  A set of five 600 s exposures was taken, with each image
slightly offset to fill in physical gaps between the MOSAIC CCDs.
The pixel scale is 0\farcs258 pixel$^{-1}$, resulting in roughly a
36$\arcmin\times36\arcmin$ field of view.

\subsection{Echelle Spectroscopy}

We obtained high-dispersion spectroscopic observations of Rosette HH\,1
with the echelle spectrograph on the Blanco 4~m telescope at the Cerro Tololo
Inter-American Observatory on 2004 January 9 and 12.  In each observation
a 79 line mm$^{-1}$ echelle grating was used.  The observations on 2004
January 9 were made in a multi-order mode, using a 226 line mm$^{-1}$
cross-disperser and a broad-band blocking filter (GG385).  The spectral
coverage is roughly 4000--7000 \AA, so that nebular lines of a range of
excitation can be examined.  In the case of the [\ion{S}{2}]
$\lambda\lambda$6717, 6731 doublet, the line ratio has been used to estimate
the electron densities within the jet.  The observations on 2004 January 12
were made in a single-order mode, using a flat mirror and a broad H$\alpha$
filter (central wavelength 6563 \AA\ with 75 \AA\ FWHM) to isolate the
order containing the H$\alpha$ and [\ion{N}{2}] $\lambda\lambda$6548, 6583
lines.  The exposure time used for both instrumental setups was 1,200 s.

For each observation the long-focus red camera was used to obtain a
reciprocal dispersion of 3.5 \AA\ mm$^{-1}$ at H$\alpha$.  The spectra
were imaged using the SITe2K \#6 CCD detector.  The 24 $\mu$m pixel size
corresponds to 0\farcs 26 pixel$^{-1}$ along the slit and $\sim$0.08 \AA\
pixel$^{-1}$ along the dispersion axis.  Both observations used a 1\farcs 6
slit oriented roughly along the jet direction, at position angles of
312$^\circ$ (multi-order) and 318$^\circ$ (single-order).  The resultant
instrumental resolution, as measured by the FWHM of the unresolved
telluric emission lines, was 0.29 \AA\ or 13 km~s$^{-1}$ at H$\alpha$.

The observations were reduced following standard procedures in the IRAF
(Ver. 2.12) software package.  This included bias correction,
flat-fielding and gain-jump removal between the chips. Cosmic-ray hits
were manually rejected from the 2D spectrograms.  Wavelength calibration
of the data was carried out based on Th-Ar lamp exposures and further
improved by comparison with night sky emission lines, which resulted in
an accuracy of $\sim$1 km~s$^{-1}$ before converting to the heliocentric
frame.

\subsection{Simultaneous Photometric and Spectroscopic Monitoring}

We have carried out a simultaneous photometric and spectroscopic
monitoring campaign of the Rosette HH1 source between 2004 December 31
and 2005 January 7. The time-series photometric observations,
unaccompanied by spectroscopy, were further extended from January 8
to January 13.  The photometric observations were made in $B$
and $R$ filters with the 0.8~m telescope of the Hsing-Hua University,
located at the Xing-Long station of the National Astronomical
Observatory of the Chinese Academy of Sciences (NAOC).
Differential photometry of the
jet-driving source was obtained through comparisons with two slightly
brighter stars in the same field at
$\alpha$(J2000) = 06$^{\rm h}32^{\rm m}15\rlap{.}{^{\rm s}}47$,
$\delta$(J2000) = $04\arcdeg55'20\farcs27$ and
$\alpha$(J2000) = $06^{\rm h}32^{\rm m}22\rlap{.}{^{\rm s}}69$,
$\delta$(J2000) = $04\arcdeg54'05\farcs31$.
The $R$ band photometry of the reference stars is found to be constant
within 0.04 mag throughout the monitoring campaign.  Many of the $B$
band exposures were affected by charge bleeding from the saturated O9.5
star HD\,46241.  These unreliable data are not presented here.

Low-resolution spectroscopy of the jet source was obtained with the
2.16~m telescope of NAOC during this monitoring campaign. Two different
spectrographs were used.  From 2004 December 31 to 2005 January 4, the
Beijing Faint Object Spectrograph and Camera (BFOSC), a copy of EFOSC
in service at the European Southern Observatory, and a thinned
back-illuminated Orbit 2k $\times$ 2k CCD were used.  The G4 grating was
employed, which gave a two-pixel resolution of 8.3~\AA.
On 2005 January 5--7, an OMR (Optomechanics Research Inc.) spectrograph
and a Tecktronix 1024 $\times$ 1024 CCD were used.  These spectroscopic
data have a higher resolution, with a 100~\AA~mm$^{-1}$ reciprocal
dispersion and a two-pixel resolution of 4.8 \AA.
Both sets of observations used a 2$\arcsec$ slit.
The spectroscopic data were reduced using standard procedures and packages
in IRAF.  The CCD reductions included bias and flat-field correction,
nebular background subtraction, and cosmic rays removal.  Wavelength
calibration was performed using He-Ar lamp exposures at both the
beginning and the end of the observations every night. Flux calibration of
each spectrum was based on observations of at least 2 of the KPNO spectral
standards \citep{Metal88} per night.

\section{Morphology and Kinematics of the Jet}

Figure 1 presents our new H$\alpha$ image of the Rosette HH1 jet.  This
high-quality image reveals that the jet does not trace back through
the exact center of the jet-driving source.  The morphology of the
jet appears to indicate episodic or nonsteady mass ejection.  A close
inspection shows a split at the end of the collimated jet, with one
branch remaining straight while the other bending north possibly
as a result of an interaction with the stellar wind of the O4 star,
HD\,46223.

In Figure 2, we present the echelle spectrograms of the Rosette
HH1 jet.  The continuum emission at the origin of each spectrogram
is from the jet-driving source.  The single-order observations
cover only the H$\alpha$ and [\ion{N}{2}] lines (the two panels
to the left in Fig.~2).  The multi-order observations detected
the H$\alpha$, H$\beta$, H$\gamma$, \ion{He}{1} $\lambda$5876,
[\ion{N}{2}] $\lambda\lambda$6548, 6583, [\ion{O}{3}]
$\lambda\lambda$4959, 5007, and [\ion{S}{2}]
$\lambda\lambda$6716, 6731 lines.
The three panels to the right in Figure 2 show the [\ion{N}{2}]
$\lambda$6583, [\ion{S}{2}] $\lambda$6731, and [\ion{O}{3}]
$\lambda$5007 lines (the brighter component of each doublet).
These lines have different thermal widths and require
different excitation energies, and thus appear different
and can be intercompared to gain physical insight.

The [\ion{N}{2}] lines have smaller thermal widths than
the H$\alpha$ line and thus resolve the velocity
structures of the HH jet and the superposed nebula more
clearly.  The [\ion{N}{2}] lines detect a prominent
irregular component at the location of the jet.  This
jet component is blue-shifted with respect to two
nebular components that have nearly uniform velocity
and surface brightness throughout the slit.
These two nebular components, at heliocentric velocities
($V_{\rm hel}$) of $\sim$13 and 40 km~s$^{-1}$, arise
from the approaching and receding sides of the Rosette
Nebula's expanding shell.
These velocities imply a systemic velocity of $V_{\rm hel}
\sim$ 27 km~s$^{-1}$ and an expansion velocity of $\sim$14
km~s$^{-1}$.
The extreme velocity of the jet reaches $V_{\rm hel}$ = $-$30
km~s$^{-1}$, which is blue-shifted from the Rosette's sytemic
velocity by 57 km~s$^{-1}$.  These results are
consistent with those reported by \citet{Metal05}.

In the single-order observation along PA = 318$^\circ$, the
[\ion{N}{2}] emission of the jet shows two bright knots and
two faint knots, with the outermost knot being the faintest.
No emission from a counterjet is detected.
The multi-order observation has a shorter slit along a slightly
different position angle, PA = 312$^\circ$, and thus shows
a slighly different velocity structure in the [\ion{N}{2}] line.
The [\ion{S}{2}] emission shows velocity structure and surface
brightness similar to those of the [\ion{N}{2}] emission.
The [\ion{O}{3}] emission, on the other hand, shows a smooth
surface brightness distribution, in contrast to the knots
seen in the other lines.

The observed velocity FWHM of the [\ion{N}{2}] line ranges
from  $\sim$16 to $\sim$23 km~s$^{-1}$, with the fainter
knots showing broader velocity widths.  These widths are not
much larger than the observed FWHM of 18-20 km~s$^{-1}$ in the
Rosette expanding shell components.  For an instrumental
FWHM of 12 km~s$^{-1}$ and a thermal width of 5.7 km~s$^{-1}$
for [\ion{N}{2}] at 10$^4$ K, the observed FWHM of the jet
implies an intrinsic turbulent FWHM of 9 to 19 km~s$^{-1}$.

\section{Stellar H$\alpha$ Profile}

One interesting spectral feature suggested by \citet{LR04}
for the jet-driving source is an inverse P Cygni profile
at the H$\alpha$ line based on a low-dispersion spectrum.
An inverse P Cygni profile, if confirmed, indicates that
material is being accreted onto the star.
Our new high-dispersion echelle observations clearly resolve
both spatially and spectrally the nebular and stellar
components of the H$\alpha$ line profile, and thus allow a
critical assessment of this suggested inverse P Cygni
profile.  As seen in Figure 2 and shown below, the bright
nebular emission from the Rosette Nebula makes it difficult
to accurately extract a clean stellar spectrum.

The nebular spectrum varies along the slit.  To assess the
nebular contribution, we have extracted seven H$\alpha$ line
profiles using 1$''$-wide windows and 0\farcs5 intervals
stepping across the stellar spectrum.  These H$\alpha$
profiles are shown in Figure 3.
The spectrum J is extracted from the jet side of the
star, and the blue-shifted jet component is clearly seen.
Seeing spreads the jet emission into the stellar spectra
S1--S4.
The contribution from the Rosette Nebula is better
represented by the spectra N1 and N2, extracted outside
the star on the side opposite to the jet.
The average of these nebular spectra is subtracted from
the four stellar spectra S1--S4.
The nebula-subtracted stellar spectra S1$'$--S4$'$, displayed
in Figure 4, show a narrow, blue-shifted emission component
and a broad, red-shifted absorption component superposed on
a continuum.

To determine the origin of the blue-shifted emission
component, we use the [\ion{N}{2}] $\lambda$6583
forbidden line that is expected only from low-density
gas, such as the Rosette Nebula and the HH1 jet.
The seven [\ion{N}{2}] line profiles extracted in a similar
manner are displayed in the right panel of Figure 3, and
they indeed show the two components from the Rosette Nebula
throughout the slit.  The [\ion{N}{2}] profiles in the four
nebula-subtracted stellar spectra, displayed in the right
panel of Figure 4, show that the nebula-subtraction
satisfactorily removes the emission from the Rosette Nebula
and that the remaining [\ion{N}{2}] emission is from the
HH1 jet.  The comparison between the nebula-subtracted
H$\alpha$ and [\ion{N}{2}] profiles suggests that the
blue-shifted H$\alpha$ emission in the stellar spectra
predominantly arises from the jet.

To obtain a clean stellar spectrum, we scale and subtract
the nebula-subtracted jet spectrum (J$'$) from the
nebula-subtracted stellar spectra (S1$'$--S4$'$) by trial
and error until the [\ion{N}{2}] emission is minimized.
To better show the stellar continuum, the spectra from each
step of this procedure are shown in Figure 5 over a larger
wavelength range.
The final clean stellar spectra S2$''$ and S3$''$ show
H$\alpha$ absorption with little or no blue-shifted emission.
The lack of strong stellar H$\alpha$ emission implies the
absence of a significant disk, making it difficult to
be associated with a jet.  This will be discussed further
in Section 7.

Our final clean  stellar H$\alpha$ line profile is quite different
from the previously reported inverse P-Cygni profile \citep{LR04}.
As we have illustrated above, the H$\alpha$ emission is dominated
by contributions from the Rosette Nebula and the HH1 jet.
These emission components can be resolved and subtracted
accurately only if high-dispersion spectra are used. The
apparent difference between the final clean stellar H$\alpha$
profiles S2$''$ and S3$''$ probably results from imperfect subtraction
of the jet component, as the [\ion{N}{2}]/H$\alpha$ ratio may vary
along the jet.  The previously reported inverse P-Cygni profile is
most likely an artifact caused by difficulties in subtracting the
nebular background and jet contribution using low-dispersion spectra.

\section{Photoionization and Physical Parameters of the Jet}

The UV radiation in the Rosette Nebula is predominantly provided
by the massive stars HD 46223 and HD 46150.
HD 46223, of spectral type O4V(f), is the hottest star in NGC 2244,
and produces Lyman photons at a rate of 10$^{49.9}$ s$^{-1}$ \citep{P73}.
It is located at 277$\arcsec$, or 2.0 pc for a distance of
1.5 kpc \citep{DM87}, from the HH1 jet source.
HD 46150 is an O5V star projected at 433$\arcsec$, or 3.1 pc,
from the jet source.  It produces ionizing photons at a rate of
10$^{49.7}$ s$^{-1}$ \citep{P73}.
The combined Lyman continuum emission from these two
exciting stars renders 1-2 orders of magnitudes higher impact
on the Rosette HH1 jet than that on similar jets discovered
in the vicinity of $\sigma$ Orionis \citep{Retal98} and the
Trapezium stars \citep{Betal00}.
The Rosette Nebula is therefore among the most extreme
environments in which photoionized jets are found. Although
immersed in a photoionized medium, the presence of highly
collimated jets strongly suggests the existence of at least a
relic disk as a sustained feed to the surviving jet. In the case
of the Rosette HH1 jet, we expect a photoevaporating disk with a
configuration resembling that of HH527 in the Orion Nebula,
as resolved by the {\it Hubble Space Telescope}
\citep{Betal00}. This could serve as a schematic
impression of the appearance of the disk-jet system.
Their configuration of the disk subject to photoevaporation
induced dissipation is believed to be similar, although the
jet associated with HH527 may be oriented at a different
direction with respect to the incident UV radiation and has a
low excitation, being located in the outskirts of the Orion
Nebula.

The electron density of the HH1 jet was derived from the [\ion{S}{2}]
doublet ratios of $\lambda$6716/$\lambda$6731 measured with
our multi-order echelle observation along the jet.
The $\lambda$6716/$\lambda$6731 ratios are 0.85$\pm$0.1 in the
jet and 1.3$\pm$0.1 in the Rosette Nebula.
The corresponding electron densities are $\sim$1000 cm$^{-3}$ in
the jet and $\le$100 cm$^{-3}$ in the backgound \ion{H}{2} region.
If the HH1 jet is indeed within the cavity of the Rosette Nebula,
the medium between the ionizing stars and the HH1 jet is hot
and ionized with a density of $\sim$0.1 H-atom cm$^{-3}$
\citep{Tetal03}.
The stellar ionizing flux at the HH1 jet would be nearly
unattenuated, at a level of $2.3\times10^{11}$ photons
cm$^{-2}$ s$^{-1}$.
For a medium of 1000 cm$^{-3}$ density, this flux can
ionize gas to a thickness of 0.37 pc.
The width of the HH1 jet (measured from Fig.\ 1) is
$\ll$1.5$''$, or $\ll$ 0.01 pc; thus the HH1 jet can be fully
photoionized by the radiation from HD 46223 and HD 46150.

Using the H$\alpha$ surface brightness of the HH1 jet,
$3.5\times10^{-5}$ ergs s$^{-1}$ cm$^{-2}$ sr$^{-1}$, and a
density of 1000 cm$^{-3}$, we find that the width (or the depth for
a cylindrical geometry) of the jet is $\sim$0.8$''$, or 0.0056 pc.
If we assume a flow velocity of 200 km s$^{-1}$, as did
\citet{Metal05}, the mass loss rate would be
$\sim1.2\times10^{-7}~M_\odot$ yr$^{-1}$. It ought to be noted
that when the disk-jet systems are exposed to photoionizing
environments, the jet production is probably no longer a dominant
role of mass loss from the circumstellar disk.  Photoionization
and dissipation of the disk then takes place, or at least
consumes the circumstellar materials at a comparable rate as the
mass ejection in the form of a jet.  If we assume that evaporated flows
associated with the dissipating disk of the jet source govern a
comparably effective mass erosion as those of the proplyds in the
Orion Nebula \citep{HO99}, then for a mean mass loss rate of 4.1 x
10$^{-7}$ $M_\odot$ yr$^{-1}$ \citep{HO99} and a disk mass of 0.006
$M_\odot$ associated with the Rosette HH1 source \citep{LR04},
the estimated photodissipation timescale of the relic disk is
$\sim$ 10${^4}$ yrs.  Given the more extreme
environment the Rosette HH1 source faces, a mass loss rate an order of
magnitude higher may be more likely and the disk dissipation time would
be reduced to $\sim10^{3}$ yrs.

%

\section{Variability of the Jet-Driving Source}

Photometric results of the jet driving source are presented in
Figure 6, which shows irregular variations around the
mean with an amplitude as large as $\sim$0.2 mag in the R band.
This amplitude of variation is about one magnitude lower than that
detected for the energy source of the Rosette HH2 jet, which
amounts to as large as 1.4 mag in R \citep{Letal07}.
Variations in the R band are primarily attributed to erratic
fluctuations of the H$\alpha$ emission of the source, which relies
on a time-variable mass accretion rate, disk inhomogeneity, or
otherwise chromospheric activity of the central YSO.
The Rosette HH1 source's low amplitude of variation
is believed to be due to a lack of circumstellar material and
subsequently a subtle mass accretion rate,
as shown by the nearly absent H$\alpha$ emission from the jet source,
although the irregular variation itself is consistent with a young
status of evolution of the central source.  We thus suggest that
the Rosette HH1 source may well represent a transient phase of
YSOs evolving rapidly from a CTTS to a WTTS by fast
photodissipation of their circumstellar disks. Based on the time
series photometric data achieved, we find no evidence of a binary
origin of the jet source, which otherwise could imply a different
mechanism of jet production. As noted in Section 2.3, the B band
observations do not give very good results, although reminiscent
irregular variations with a comparable magnitude of up to
$\sim$0.25 mag are indeed indicated.

The spectral monitoring observations are not very useful for the
H$\alpha$ line profile because of the difficulty in background
nebular subtraction, as discussed in Section 4. Nevertheless, the
overall spectral characteristics of the star can be determined. We
find that the spectral type appears to vary between F8V and F9V
during the period of observations. This spectral change may be
related to photo-erosion of the rotational disk with an
inhomogeneous configuration.

\section{Evidence for Fast Disk Dissipation and a Young Stellar Age?}

Being immersed in the fierce UV radiation field of the Rosette,
the optical jets associated with YSOs indicate either a
jet production timescale of as long as 1-2 Myr, comparable to the
evolutionary age of the main cluster NGC 2244, or that the YSOs
have a much younger age and the cocoons associated with their
protostars had, in some way, been successfully shielded from the
strong ionization fields.

\citet{Li05} investigated the YSOs with near infrared excesses, an
indicator of the existence of circumstellar disks, of the young
open cluster NGC 2244 based on the 2MASS database. The jet-driving
sources in the Rosette, however, show infrared colors commensurate
with those of WTTSs, which have spectral energy distributions
indistinguishable from main-sequence dwarfs.
See the color-color and color-magnitude diagrams in Figures 4 and
7 of \citet{Li05}. This, along
with the fact that none of the Rosette jet sources and their only
rivals found near $\sigma$ Orionis were detected by IRAS, suggests
a lack of circumstellar material as compared to conventional YSOs
driving outflows. This is in agreement with the estimated mass
of 0.006 $M_\odot$ for the relic disk associated with the Rosette HH1
source \citep{LR04}, far below the typical value of $\sim$0.1
$M_\odot$ around CTTS.  Given the emerging nature of the young
open cluster with a turnoff age of 1.9 Myr \citep{PS02}, fast disk
dissipation is suggested. \citet{LR04} suggest that this provides
indirect observational evidence for the formation of isolated BDs
and free-floating giant planets, as discovered in Orion by
\citet{Zetal00}, by UV dissipation of unshielded protostellar
systems. This can be very important to our understanding of the
formation of such sub-stellar and planetary mass objects,
particularly in regions of massive star formation. Such UV
dissipation could, on the other hand, impose strong effects on the
formation of and hence the search for extra-solar planets around
low-mass stars, the circumstellar disks of which could otherwise
be potential sites of terrestrial planet formation. This
alternatively introduces a viable solution to the long puzzle of
how WTTS were formed as a consequence of fast CTTS evolution and
the rapid dissipation of circumstellar disks under particular
forming conditions near massive OB stars or in cluster
environments.

The spatial distribution of the extreme jets with respect to the
dozens of exciting OB stars of the spectacular \ion{H}{2} region
is presented in Figure 7, superimposed on which is the relic shell
structure as delineated by the apparent congregation of excessive
emission sources in the near infrared \citep{Li05}. This suggests
the existence of a former working interface layer of the
\ion{H}{2} region with its ambient molecular clouds. The projected
location of the energy source of the Rosette HH1 jet near the
relic arc provides evidence of a triggered origin of its formation
in or near the swept-up layer. The Rosette HH2 source has a
similar radial distance from the statistical center of NGC 2244
\citep{Li05} and introduces a similar origin.  In this scenario,
the jet sources should have a much younger age than the main
cluster NGC 2244. Molecular gas and dust in the shell could have
played an important role in shielding new generation protostellar
objects from the harsh UV evaporation and ionization from the
massive OB stars. It is therefore reasonable to infer that the jet
sources have been directly exposed to the harsh photoionization
fields recently.


\section{Large-Scale High-Excitation Structures in the Rosette Nebula}

\citet{MW86} first noted the existence of ionized knots and
filaments in the southeastern quadrant of the Rosette's central
cavity.  The Rosette HH jets are also located in this region.  All
are prominent features in narrowband [\ion{O}{3}] images,
indicating a high excitation. Among these, knot C shows
high-velocity components and is in association with a high
excitation bow-shock at its tip \citep{CM95,Cetal98}. This feature
is believed to be a HH flow, though no apparent energy source has
yet been identified.

However, a giant shock-like structure to the west of the Rosette
HH1 jet can be easily identified (Figure 8). At a distance of
$\sim$1.5 kpc, its large-scale appearance and lack of a potential
exciting source seem to exclude the possibility of a Herbig-Haro
origin. The preferential distribution of these high-excitation
structures to the southeast edge of the \ion{H}{2} region suggests
a possible association with the large Monoceros Loop supernova
remnant (SNR) projected to the northeast of the Rosette Nebula.
While there is no morphological evidence for dynamical
interactions between these two objects, it is possible that some
of the high-excitation structures in the Rosette Nebula are caused
by the Monoceros Loop SNR's ballistic ejecta that proceeds ahead
of the SNR shock front.  To test this scenario, proper motion or
abundance measurements of the high-excitation structures in the
Rosette Nebula are needed.

Alternatively, we propose that these structures are globules or
former dust pillars, similar to those around the working surface
of the \ion{H}{2} region, that have been overrun by the ionization
front and are now in the process of photodissipation, as is the fate
of the HH jets in this region.  At least one high-excitation structure
is likely associated with a neutral cometary knot (see Figure 8), the
tip of which is highly ionized and has an appearance resembling those
in the Orion Nebula \citep{Betal00}, but with a physical size
roughly 4 times larger.

\section{Summary}

We present follow-up high-qaulity imaging and echelle spectroscopic
observations of the Rosette HH1 jet.  The high angular and
spectral resolution allow us to determine accurately the
contributions from the \ion{H}{2} region, jet, and star.
The expansion of the \ion{H}{2} region and the kinematics
of the jet are consistent with the previous measurements
by \citet{Metal05}.  Using the [\ion{S}{2}] doublet ratios,
we further determined the electron density of the jet,
$\sim$1000 cm$^{-3}$.

With a careful subtraction of the nebular
and jet components, we find the stellar H$\alpha$ line is
dominated by a broad absorption profile with little or no
emission component, indicating a lack of substantial
circumstellar material.
The circumstellar material has most likely been photo-evaporated
by the strong UV radiation field in the Rosette Nebula.
The evaporation time scale is 10$^3$ -- 10$^4$ yr.
The Rosette HH1 jet source provides evidence for an
accelerated evolution from a CTTS to a WTTS due to the
strong UV radiation field; therefore, both CTTSs and WTTSs
can be spatially mixed in regions with massive star formation.

Finally, we suggest that the giant high-excitation structures residing
at the center of the Rosette Nebula may be globules or former dust
pillars in the midst of UV dissipation.  Further observations of
the nebular kinematics are needed to determine whether these are
dissipating interstellar structures or related to the supernova
ejecta associated with the Monoceros Loop SNR.

\acknowledgements We greatly appreciate the helpful comments and
suggestions from the referee of the paper, John Meaburn. Thanks to
the team working with the Hsing-Hua 80cm telescope for their help
on coordinating the photometric observations. This project is
supported by the National Natural Science Foundation of China
through grant No.10503006.

\bibliographystyle{aa}

\clearpage

\figcaption[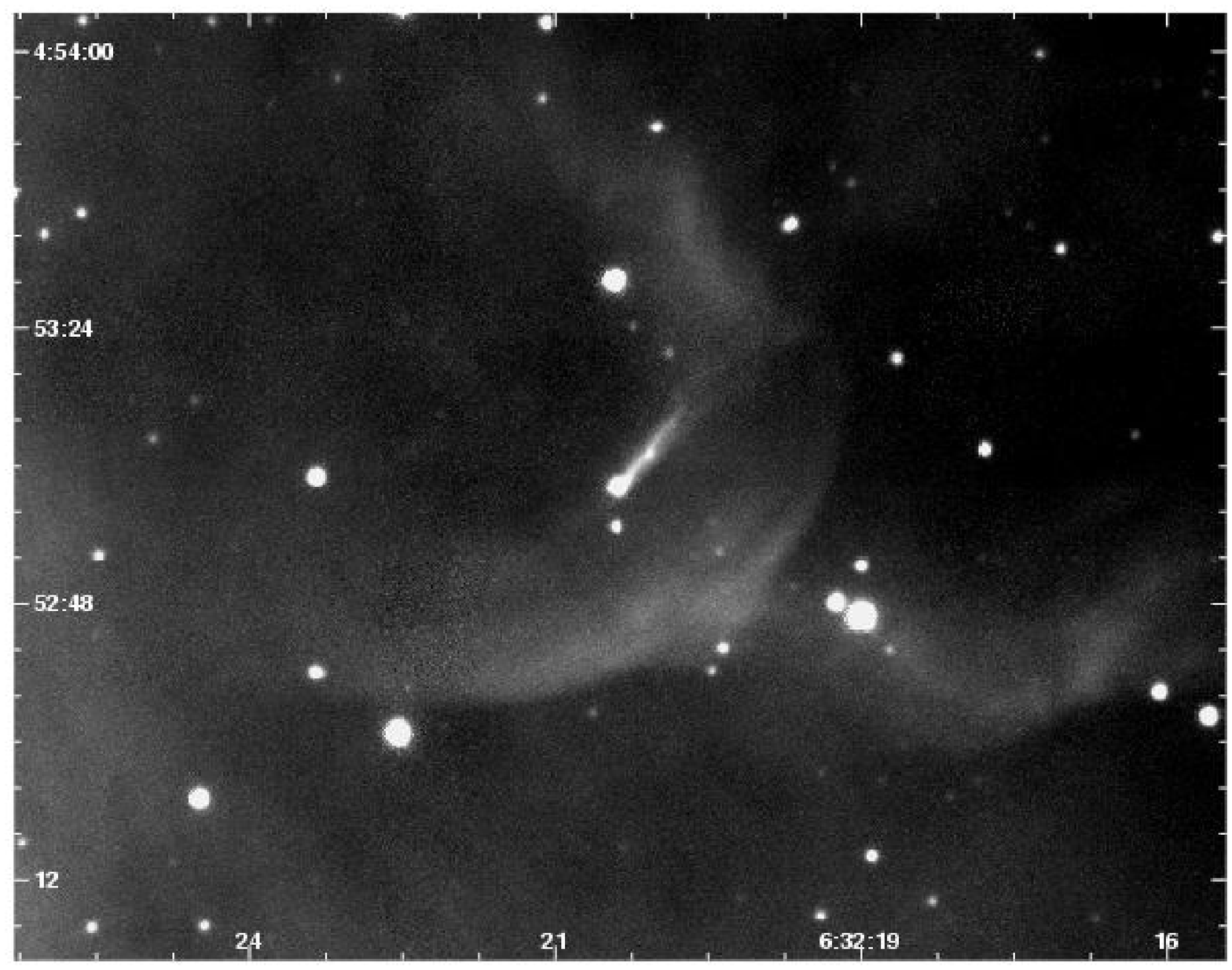]{H$\alpha$ image of the Rosette HH1 jet obtained
with the KPNO 4 m telescope.  North is up, and east is to the left.}

\figcaption[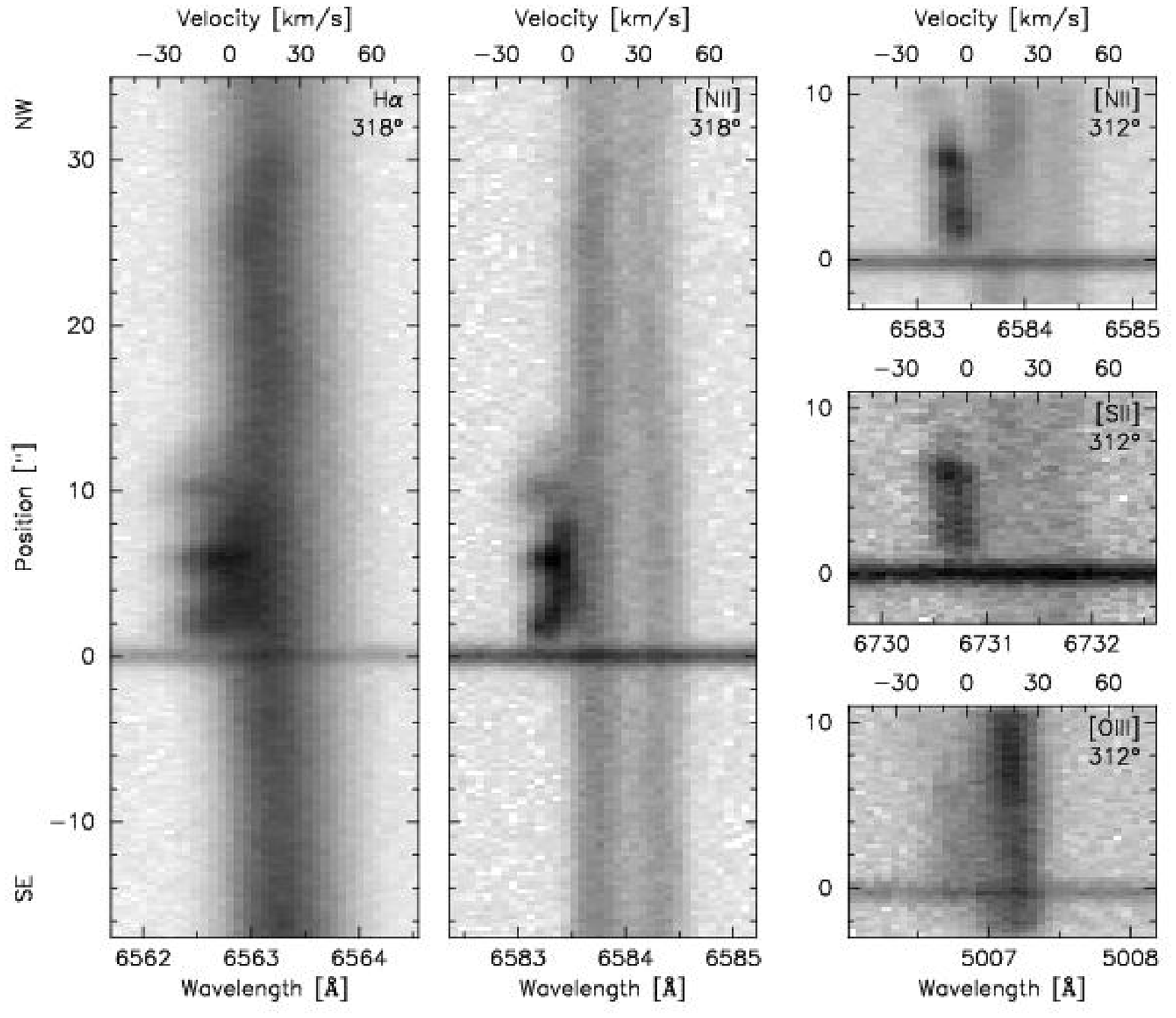]{Echelle spectrograms of the Rosette HH1 jet.
Single-order echelle spectrograms of the jet system covering only
the H$\alpha$ and [\ion{N}{2}] $\lambda$6583 lines are presented in the
left panels. The continuum emission at $Y = 0''$ of each
spectrogram is from the jet-driving source. The multi-order
observations of the [\ion{N}{2}] $\lambda$6583,
[\ion{S}{2}] $\lambda$6731 and [\ion{O}{3}] $\lambda$5007
lines are shown in the right panels.}

\figcaption[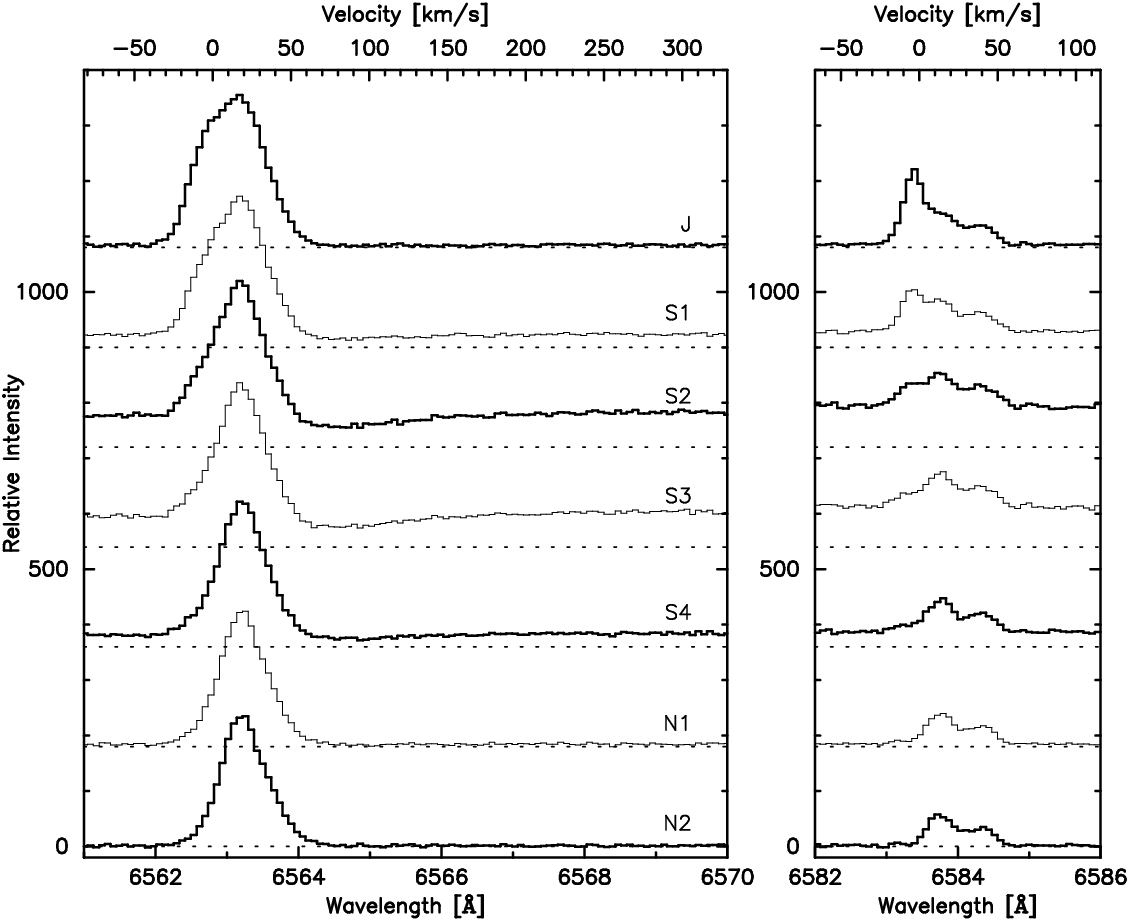]{Line profiles of H$\alpha$ (left panel) and
[\ion{N}{2}] $\lambda$6583 (right panel) extracted at positions
across the stellar spectrum, using 1$\arcsec$ windows and
0$\arcsec$.5 steps.  These profiles show contributions from
the stellar continuum, jet emission, and background nebular
emission.}

\figcaption[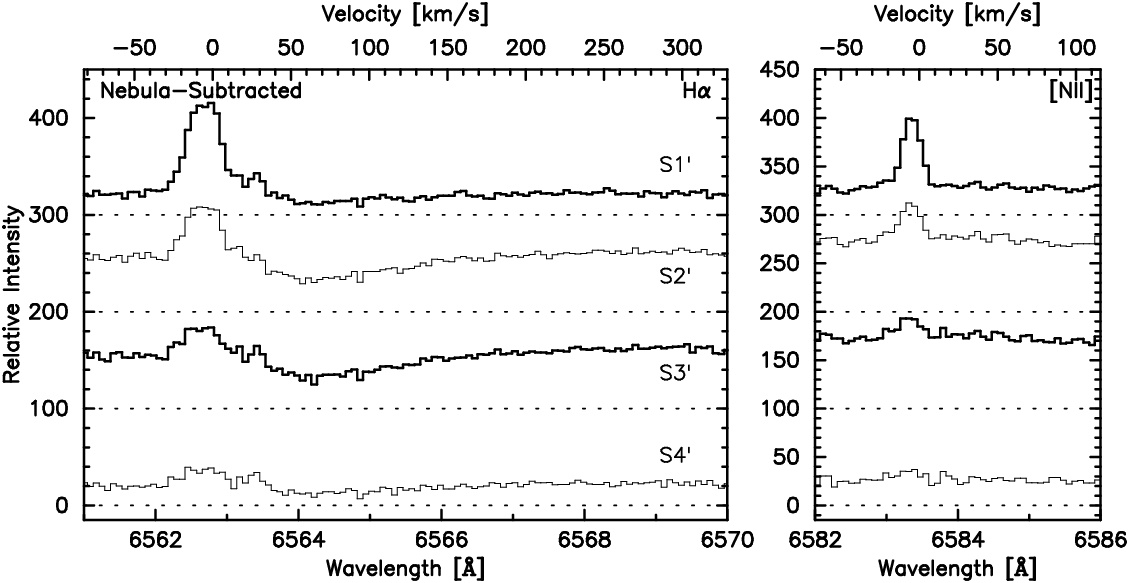]{Nebula-subtracted stellar spectra around
H$\alpha$ (left panel) and [\ion{N}{2}] $\lambda$6583 (right panel,
serving as a quality control of the nebula subtraction). Note the
narrow, blue-shifted emission component and the broad, red-shifted
absorption associated with H$\alpha$.}

\figcaption[f5.ps]{The net stellar spectrum with a larger
wavelength coverage. Top panel: nebula-subtracted stellar
spectrum. Middle panel: the clean jet spectrum after background
subtraction. Bottom panel: net stellar spectrum after the removal
of both the nebular background and the jet contribution.}

\figcaption[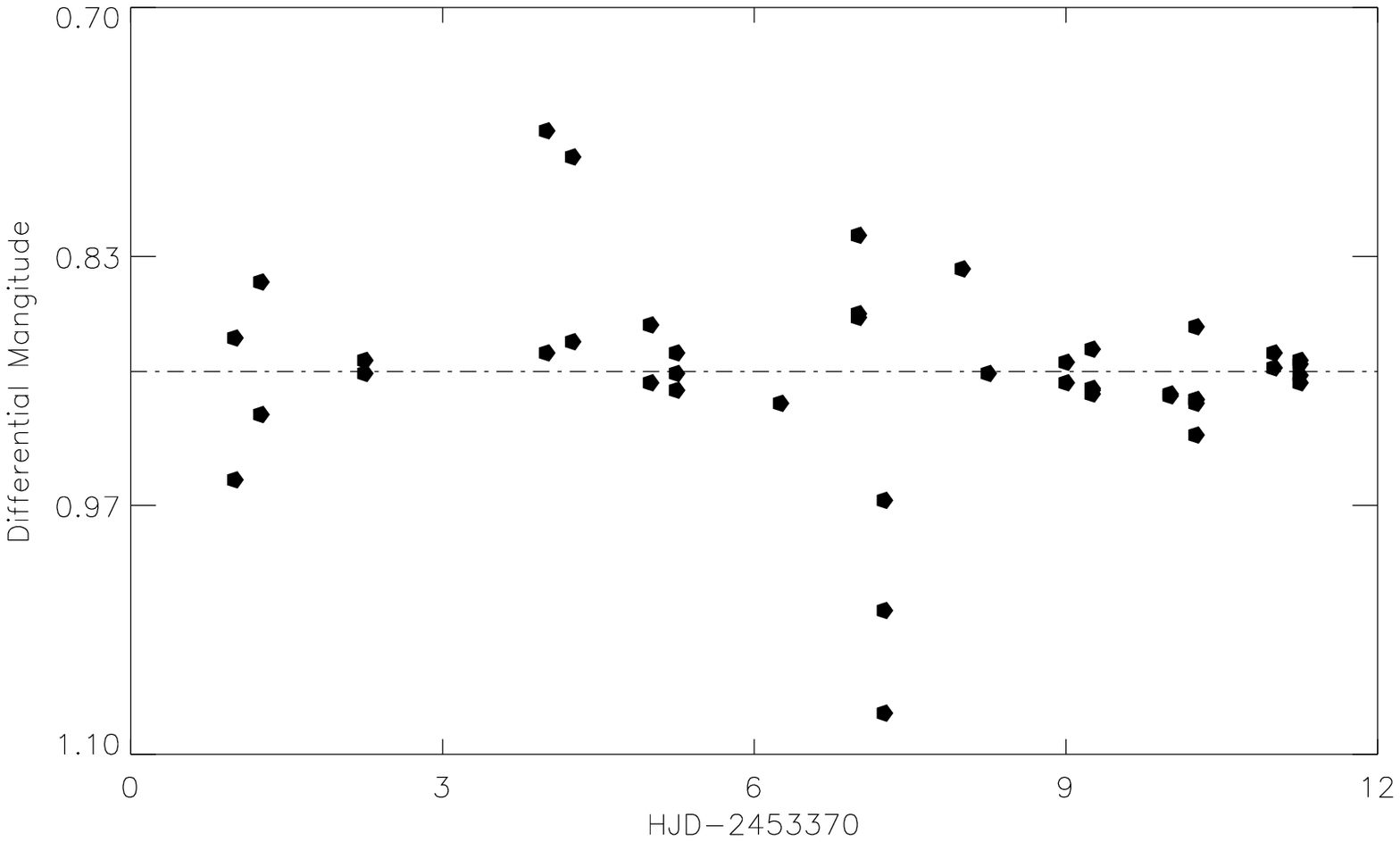]{Photometric variations of the jet source in the
R band.  Note the irregular variations at low amplitudes, which
are detected on timescales of minutes to hours around a fitted
mean level (dot-dashed line).}

\figcaption[f7.ps]{Spatial distribution of the Rosette jets with
respect to the ionizing massive OB stars that excavated the
\ion{H}{2} region (Townsley et al. 2003). The spectral types of
the OB stars are also marked on the DSS (Digital Sky Survey) R
band image of Rosette.}

\figcaption[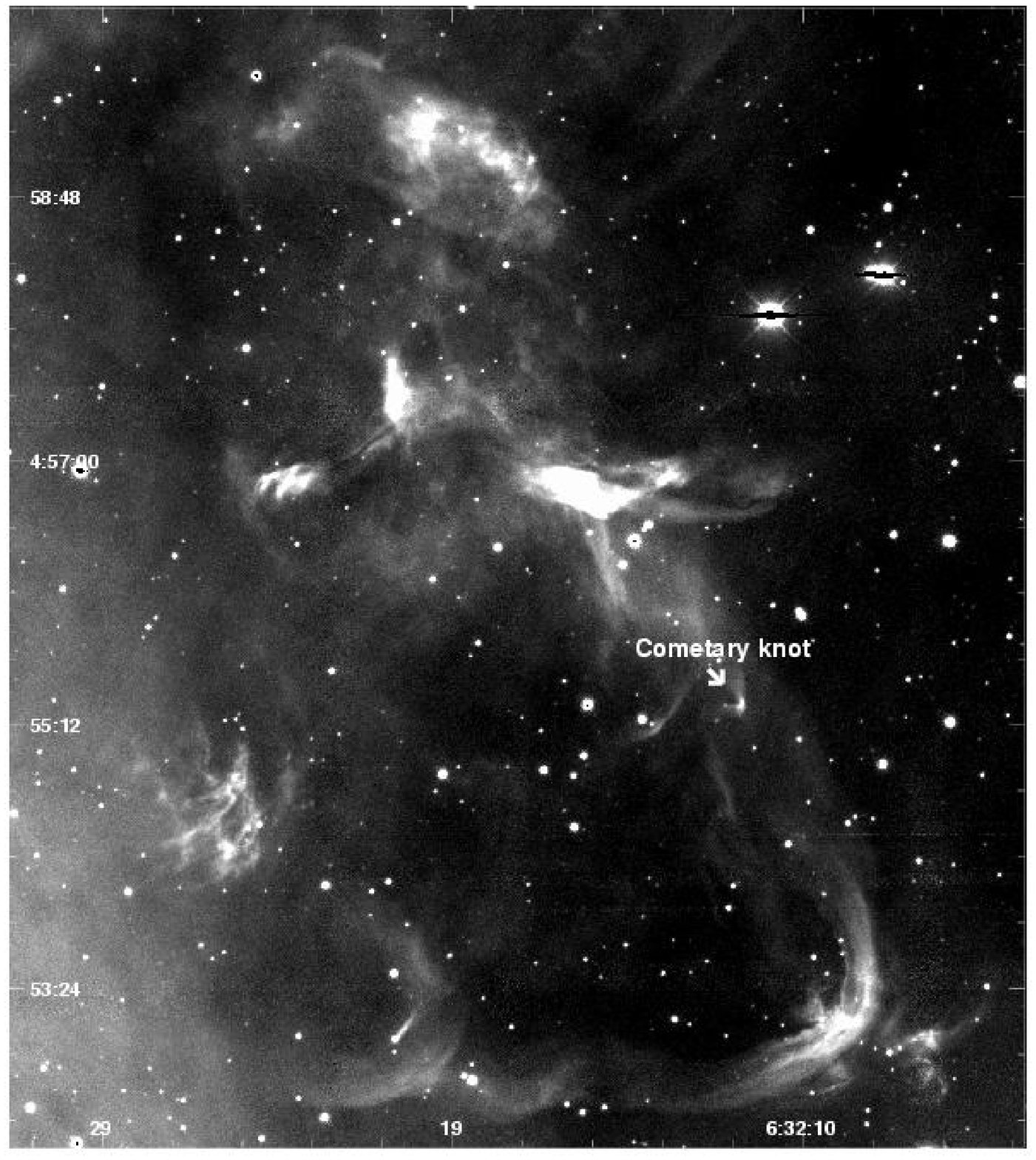]{Large high-excitation shock structures
in the Rosette Nebula.  Note the presence of at least one cometary
knot among these seemingly inter-related structures. The tip of
the neutral knot is clearly highly ionized and has an appearance
resembling those identified in the Orion Nebula \citep{Betal00},
but $\sim$4 times larger.}

\clearpage

\begin{figure}
\plotone{f1.ps}
\centerline{f1.ps}
\end{figure}
\clearpage

\begin{figure}
\plotone{f2.ps}
\centerline{f2.ps}
\end{figure}
\clearpage

\begin{figure}
\plotone{f3.ps}
\centerline{f3.ps}
\end{figure}
\clearpage

\begin{figure}
\plotone{f4.ps}
\centerline{f4.ps}
\end{figure}
\clearpage

\begin{figure}
\plotone{f5.ps}
\centerline{f5.ps}
\end{figure}
\clearpage

\begin{figure}
\plotone{f6.ps}
\centerline{f6.ps}
\end{figure}
\clearpage

\begin{figure}
\plotone{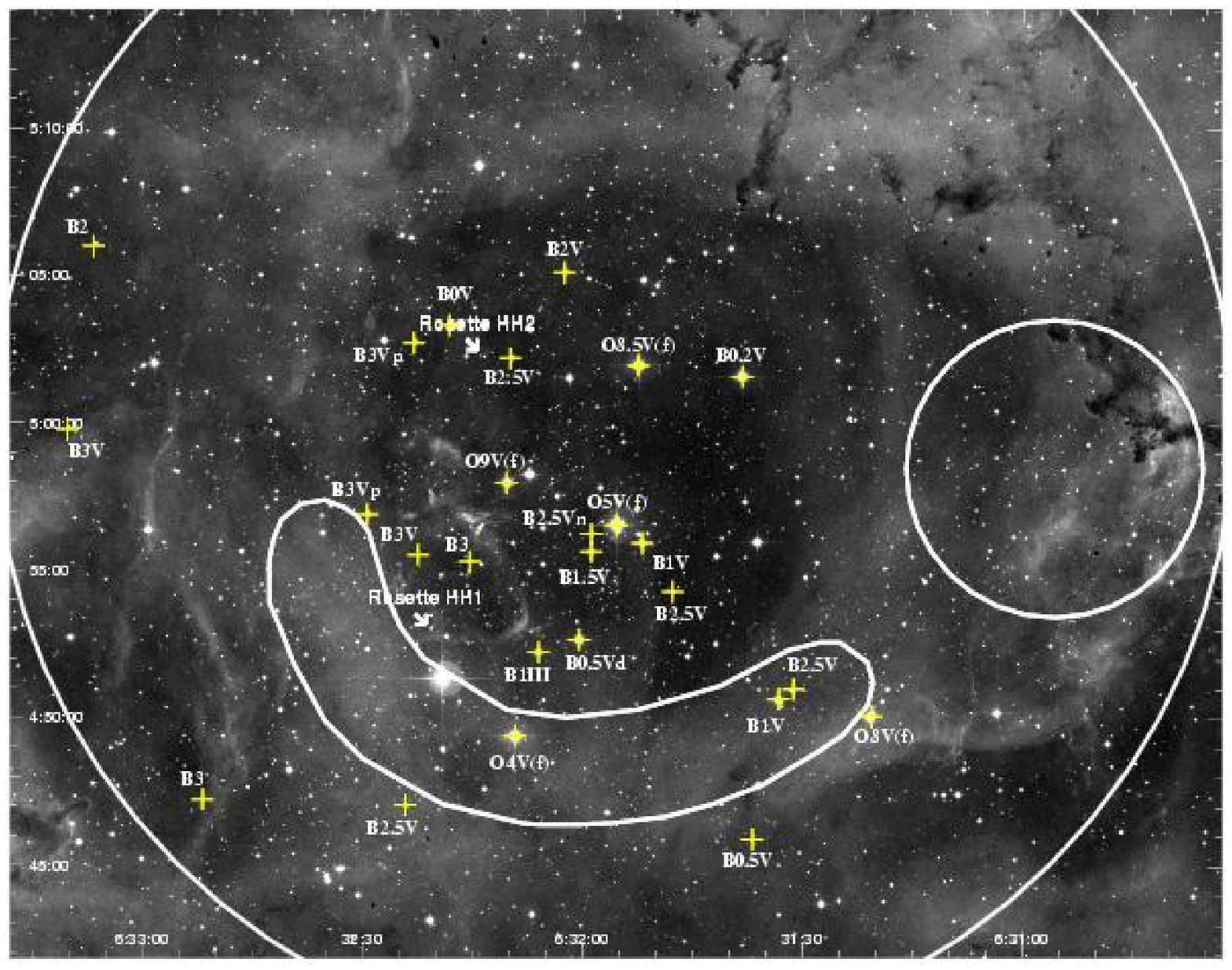}
\centerline{f7\_color.ps}
\end{figure}
\clearpage

\begin{figure}
\plotone{f8.ps}
\centerline{f8.ps}
\end{figure}

\end{document}